\documentclass[12pt,a4paper]{revtex4}
\usepackage{amsmath}  % Para reconocer s\'{\i}mbolos AMS \'{o} AMS NOT
\usepackage{amssymb}  % Para reconocer s\'{\i}mbolos raros como \nRightarrow
\usepackage{dsfont}   % Para reconocer s\'{\i}mbolos de conjuntos (N,Z,R,C,I,...)

\begin{document}
\title{On position, momentum and their correlation}
\author{Alberto C. de la Torre}\email{delatorre@mdp.edu.ar}
% \author{NNNN}\email{nnnn@mdp.edu.ar}
\affiliation{ IFIMAR - (CONICET-UNMDP) \\ Departamento de F\'{\i}sica -
 Facultad de Ciencias Exactas y Naturales -
 Universidad Nacional de Mar del Plata\\
 Funes 3350, 7600 Mar del Plata, Argentina.}
\begin{abstract}
Position and momentum observables are considered and their
correlation is studied for the simplest quantum system of a free
particle moving in one dimension. The algebra and the eigenvalue
problem for the correlation observable is presented and its possible
relevance for the solution of the Pauli problem is analysed. The
correlation provides a simple explanation of the shrinking and
spreading of wave packets in an interpretation of quantum mechanics
based in an ontology suggested by quantum field theory. Several
properties and speculations concerning position-momentum
correlations are mentioned.
%\\ Published in  \textbf{30},  467–475, (2009) arXiv: 0808.3083
\end{abstract}
\maketitle
%\date{\today}
\section{INTRODUCTION}
The relation between position and momentum in quantum mechanic has
many  features unexpected from the classical point of view. The
simplest quantum system consisting in one structureless free
particle moving in one dimensional space has position and momentum
as unique relevant observables. Although this is the simplest system
we may think of, it has sufficient weirdness to justify the claim
that, from the point of view of the foundations of quantum
mechanics, we can say that we do not completely understand it. In
this work we will present several arguments supporting this claim
and we will give a detailed analysis of the correlation observable
for position and momentum that may bring some more knowledge on the
system in the attempt to reach a better understanding.

\section{QUANTUM MECHANIC RECIPE}
According to standard quantum mechanics, in order to describe one
free structureless particle moving in one dimension we define a
(Rigged) Hilbert space $\mathcal{H}$ whose elements
$\Psi\in\mathcal{H}$ represent the state that encodes all
information about the system. Position and momentum are represented
by hermitian operators $X$ and $P$ in $\mathcal{H}$ such that their
associated bases, $\{\varphi_{x}\}$ and $\{\phi_{p}\}$, are mutually
unbiased. Their internal product is
$\langle\varphi_{x},\phi_{p}\rangle=
\frac{1}{\sqrt{2\pi}}\exp(\frac{i}{\hbar}px)$. $P$ and $X$ are,
respectively, the generators of translations $X\rightarrow
X+a\mathbb{I}$ and impulsions $P\rightarrow X+g\mathbb{I}$ and
therefore (if the dimension of $\mathcal{H}$ is infinite) their
commutator is $[X,P]=i\hbar$. The quantum mechanical prediction is
that the measurement of position or momentum in an ensemble of
systems in the state $\Psi$ will be distributed according to the
densities $\rho (x)$ and $ \varpi (p)$ given by
\begin{eqnarray}
% \nonumber to remove numbering (before each equation)
    \rho (x) &=&  |\langle\varphi_{x},\Psi\rangle|^{2} \ ,\\
  \varpi (p) &=&  |\langle\phi_{p},\Psi\rangle|^{2} \ ,
\end{eqnarray}
and similarly, the prediction for any observable $C(X,P)$ is given
in terms of its eigenvectors $\{\eta_{c}\}$ by the density
\begin{equation}\label{poi}
   \sigma(c) = |\langle\eta_{c},\Psi\rangle|^{2}\ .
\end{equation}
The state $\Psi$ of the system is determined after the measurement
of any observable $A(X,P)$ with result $\alpha$ by the eigenvalue
equation $A(X,P)\Psi=\alpha\Psi$ (for simplicity we assume pure
states and nondegeneracy). The time evolution of the state, and
therefore of any distribution, is controlled by a unitary operator
$U_{t}$ such that $\Psi_{t}=U_{t}\Psi_{0}$ with $U_{t}=
\exp(-\frac{i}{\hbar}tH)$ where $H$ is the hamiltonian ($P^{2}/2m$
in this case). That's all.

\section{EXISTENTIAL WEIGHT}
Does the recipe above provide sufficient explanation for the
complete understanding of the system? There are arguments to deny
this. Although many physicists would feel satisfied because every
prediction can be tested in a laboratory measurement, there are
questions concerning the reality of the system that do not refer to
laboratory measurements but, for many of us, their answer is
necessary for a complete understanding of the system. What is the
nature of the distributions $\rho (x),\ \varpi (p)$ and $
\sigma(c)$? The standard answer that ``they are probability
distributions'' is not really satisfactory\cite{dlt1}. Indeed, if
position and momentum are random variables with their corresponding
probability distributions given by $\rho (x)$ and $\varpi (p)$ then
the well established theory of random variables provides the
probability distribution for any function $C(X,P)$ that turns out to
be different from the quantum mechanical prediction given in
Eq.[\ref{poi}]. Consequently $\rho (x)$ and $\varpi (p)$ are not the
probability density functions of some stochastic process for a
particle diffusing in space with some random velocity. Even though
$\rho (x),\ \varpi (p)$ and $ \sigma(c)$ are related to frequency
measurements, as if they were probabilities, strictly speaking they
are not and it would be better to use another name to denote them.
The term \emph{existential weight} has been proposed\cite{dlt1}
although the misnomer ``probability distribution'' appears to be
irreversibly installed in quantum mechanics.

\section{ONTOLOGICAL INDEFINITENESS}
Let us consider now one of these existential weights, say $\rho
(x)$. If we make an experiment to detect the position of the
particle in an ensemble of systems we will find that the eigenvalues
of the position observable are distributed according to $\rho (x)$.
Now, what is the nature of this distribution? Should we think that
the particle has a definite position, the \emph{putative
value}\cite{ish}, that can not be determined by quantum mechanics
and the best that the theory can provide is the distribution $\rho
(x)$ of the observed values? In this case $\rho (x)$ reflects our
ignorance of the reality (gnoseological interpretation) and we can
think that the experiment reveals a pre-existent value of the
observable. On the contrary, we may think that the position
observable doesn't have a precise value and $\rho (x)$ represents
this inherent indefiniteness in the observable (ontological
interpretation). In this case we don't have a pre-existent value for
the observable and the experiment creates the observed result. The
gnoseological interpretation appears easier to accept, however there
are very strong arguments against it: the Bell-Kochen-Specker
theorem\cite{bell,koch} applied to the position
observable\cite{dlt2} shows that the existence of \emph{context
independent} putative values for  position is in contradiction with
quantum mechanics. There is however a stronger argument: the
violation of Bell´s inequalities\cite{bell1,exp,chsh} imply that the
existence of context independent putative values is in
contradiction, not only with quantum mechanics, but also with
empirical reality.

Context independence, means that the putative value should not
depend on the value taken by other \emph{commuting} observables, for
instance, the position or momentum of another noninteracting
particle located far away or any other commuting observable that the
theorist may think of. Although \emph{context dependent} putative
values are not excluded, their existence appears suspicious and
difficult to accept and therefore many experts in the foundations of
quantum mechanics adopt the ontological interpretation of the
distributions even though it is not a logical necessity.

\section{POSITION AND MOMENTUM DEFINITION }
If  position and momentum of a particle do not assume exact values
but are instead diffuse by nature, as the ontological interpretation
of $\rho (x)$ and $\varpi (p)$ suggests, then we must review the
intuitive understanding of the relation between position and
momentum. Of course, in this diffuse position case we can not define
the velocity as the time derivative of position because there is no
such a position. Therefore the momentum definition of elementary
mechanics $P=mV$ is not acceptable. Anyway, this definition is
abandoned also in classical physics because it is incorrect for a
charged particle in the presence of an external electromagnetic
field (furthermore it results inadequate in special relativity). The
standard way in quantum mechanics is to define momentum as the
observable whose associated operator has the commutation relation
$[X,P]=i\hbar$ with position and that can also be understood as
generator of translations because we can prove that, for the
observable whose operator $F(X)$ is a function of position (that can
be expanded in a power series) we have
\begin{equation}\label{commutPderivX}
   [F(X),P]=i\hbar\frac{dF(X)}{dX}\ ,
\end{equation}
and therefore $U^{\dag}_{a}F(X)U_{a}=F(X+a)$, where
$U_{a}=\exp(-\frac{i}{\hbar}aP)$. Similarly, the commutator with $X$
acts as a derivative with respect to $P$
\begin{equation}\label{commutXderivP}
   [G(P),X]=-i\hbar\frac{dG(P)}{dP}\ .
\end{equation}
These relations can be generalized, \emph{only in some cases}, to
functions of $X$ and $P$ taking partial derivatives in the right
hand side. This is not always possible because the chain rule of
derivatives becomes ambiguous due to the noncommutativity of $X$ and
$P$. The generalization should be used with extreme care only for
functions like $\sum a_{kr}X^{k}P^{r}$ where the partial derivatives
are unambiguous. For instance, a careless application of
Eq.(\ref{commutPderivX}) for $F(X,P)=e^{\frac{-i}{2\hbar}(XP+PX)}$
using the chain rule leads to the obviously wrong result:
$Pe^{\frac{-i}{2\hbar}(XP+PX)}=0$.

Notice however that this definition of momentum relies on the formal
aspects of the recipe of quantum mechanics and has lost the strong
connection found in classical physics where momentum is related to
matter in motion (as is suggested in $P=mV$). The lost connection
between position and momentum observable in quantum mechanics
results in that, even if we know the time evolution of position,
that is, if we know $\rho (x,t)$, we can not derive from it the
momentum distribution $\varpi (p)$. To prove this,
consider\begin{eqnarray}
% \nonumber to remove numbering (before each equation)
   \varpi (p) &=& \langle\Psi,\phi_{p}\rangle\langle\phi_{p},\Psi\rangle
   = \langle\Psi_{t},\phi_{p}\rangle\langle\phi_{p},\Psi_{t}\rangle\\
   &=& \int dx\ dx'\ \langle\Psi_{t},\varphi_{x}\rangle\langle\varphi_{x},
   \phi_{p}\rangle\langle\phi_{p},\varphi_{x'}\rangle
   \langle\varphi_{x'},\Psi_{t}\rangle  \\
 &=& \int dx\ dx'\ f(x,x')\langle
 \varphi_{x'},\Psi_{t}\rangle\langle\Psi_{t},\varphi_{x }\rangle  \ ,
\end{eqnarray}
where $f(x,x')$ is a known function but of
$\langle\varphi_{x'},\Psi_{t}\rangle\langle\Psi_{t},\varphi_{x}\rangle
$ we know only the ``diagonal'' terms given by $\rho (x,t)$.
Viceversa, in a similar way one can easily prove that the knowledge
of the momentum distribution $\varpi (p)$ and of an initial position
distribution $\rho (x)$ are not sufficient in order to calculate the
time evolution of the position distribution $\rho (x,t)$. \emph{The
complete information on position and movement of the system encoded
in the existential weights $\rho (x,t)$ and $\varpi (p)$ is not
sufficient for a complete determination of the state of the system
$\Psi$.} This fact, surprising from the classical point of view, was
first recognized by Pauli\cite{pau} and triggered an intense
investigation on the necessary and sufficient information needed for
an unambiguous state
determination\cite{wei1,wei2,qinf,unbiaBas1,unbiaBas2,dlt4}.

The existence of \emph{Pauli partners} states, that is, different
states $\psi\neq\phi$ having the same distributions for position
$\rho (x)$ and momentum $\varpi (p)$, implies that, in general,
there exists no functional $\mathcal{C}$ such that the distribution
of the eigenvalues of some observable $C(X,P)$ is given by
$\mathcal{C}(\rho (x),\varpi (p))$. Besides the knowledge of
position and movement of the system we need some additional
information concerning some function of position and movement (the
correlation perhaps) in order to determine the state of the system.
What is the cause and physical meaning of this additional function
that contributes to fix the state of the system? Is there something
besides position and momentum in the ontology of the system that is
described by such a function? is this some consequence of an unknown
geometrical space-time structure? These are questions indicative of
our lack of understanding of quantum mechanics even at the
elementary level of this simplest physical system.

\section{POSITION-MOMENTUM CORRELATION OBSERVABLE}
We can conclude from the arguments of the previous sections that,
besides position and momentum, we need another observable to provide
additional information on the system that may render possible the
determination of the state of the system without the possible
ambiguity of the Pauli partners.

One could first think that the best observable to provide the
additional information should be an observable whose basis is
unbiased to the bases of position and momentum. For instance, every
linear combination $\alpha X+\beta P$ has a basis unbiased to
$\{\varphi_{x}\}$ and $\{\phi_{p}\}$. This ``unbiased'' observable
would bring information with the highest independence from position
and momentum and therefore we may think that it is an optimum
choice. However this is not true. The Pauli partners ambiguity is
not avoided by this choice. One can prove that there is an infinite
number of states having the same existential weight for position,
momentum, and the third unbiased observable. This follows from the
fact that the three bases associated to the three observables are
mutually unbiased. However (if the Hilbert space is infinite
dimensional) there exist a fourth, and infinite many, other bases
unbiased to the previous three. Every element of these bases is an
example of a state with identical (uniform) distributions for all
three observables. That is, all the basis elements are Pauli
partners. Therefore, if we hope to resolve the Pauli partners
ambiguity, besides position and momentum we should include a third
observable whose basis is not unbiased with position and momentum.
One observable with this requirement is the correlation, defined as
\begin{equation}\label{Corr}
    C=\frac{1}{2}(XP+PX)\ .
\end{equation}

Besides providing a \emph{possibility} to solve the Pauli problem,
this observable is interesting in itself and therefore we will now
study its properties. The commutation relations with $X$ and $P$ are
\begin{equation}\label{ComRel}
   [X,C]= i\hbar X\ \hbox{ and }\ [P,C]= -i\hbar P \ ,
\end{equation}
and by induction we can prove that
\begin{equation}\label{ComRel1}
   [X^{n},C]= i\hbar nX^{n}\ \hbox{ and }\ [P^{n},C]= -i\hbar nP^{n} \
   .
\end{equation}
From these Eqs.(\ref{ComRel1}) and using Leibniz rule for the
commutator of products we get
\begin{equation}\label{ComRel3}
   [X^{r}P^{s},C]= i\hbar (r-s)X^{r}P^{s}\ \hbox{ and }
   \ [P^{s}X^{r},C]=  i\hbar (r-s)P^{s}X^{r} \
   ,
\end{equation}
and therefore, for any hermitian operator of the type
$D_{rs}=X^{r}P^{s}+P^{s}X^{r}$ we have
\begin{equation}\label{ComRel4}
   [D_{rs},C]= i\hbar (r-s)D_{rs}\ .
\end{equation}
Notice that all these commutation relations above are of the type
$[B,C]=ikB$ where $k$ is a real constant. This relation will be
relevant later when we study the eigenvectors of $C$.

One interesting thing is to \emph{try} to determine an operator
$A(X,P)$ that has, with the correlation operator $C(X,P)$, the
commutation relation equal to the corresponding commutator of
position and momentum, that is, $[A,C]= i\hbar $. Physically, we are
looking for an observable that could act as a \emph{generator of
correlations}. One can prove that such an operator $A(X,P)$ can not
be expanded in a power series $\sum a_{kr} X^{k}P^{r}$ (notice that
any power series can be brought to this ``normal order'' with all
powers of $X$ at the left of all powers of $P$). In order to prove
this we use the commutation relations in Eq.(\ref{ComRel3}) and we
can see that there exists no choice of the coefficients $a_{kr}$
that satisfy the commutation relation $[A,C]=i\hbar$.

The eigenvectors of $A(X,P)$ (if they exist) and $C(X,P)$ would
build two mutually unbiased bases and these two observables could be
chosen as a pair of canonical conjugate coordinates for the
description of the system. This choice is related to the canonical
transformation of classical mechanics where the coordinates $(x,p)$
are transformed to $a(x,p)$ and $c(x,p)$ in a way to preserve the
Poisson brackets, that is $\{x,p\}=\{a,c\}=1$. If we take
$c(x,p)=xp$, then the conjugate coordinate is $a(x,p)=\frac{1}{2}
\ln(\frac{x}{p})$. Following this suggestion we can see that the
operator
\begin{equation}\label{AcanconjtoC}
 A=\frac{1}{2}(\ln X - \ln P)\ ,
\end{equation}
at least \emph{formally}, has the wanted commutation relation. To
prove this we use Eqs.(\ref{commutPderivX}) and
(\ref{commutXderivP}) in order to obtain $[\ln X,P]=i\hbar/X$ and
$[\ln P,X]=-i\hbar/P$. With more mathematical rigour it is not clear
that such an operator exists. Furthermore, the physical meaning of
an observable such as $\ln X$, undefined for negative values of
position, is unclear, leaving alone what would be the mysterious
physical procedure to measure $A$. The question of the existence of
a generator of correlations is open.

We come now to the question of the existence of the eigenvectors of
the correlation operator $C$, that is, to determine the basis
$\{\eta_{c}\}$ and the real numbers $c$ such that
\begin{equation}\label{eigenvec}
   C\eta_{c}=c\eta_{c}\ .
\end{equation}
Strictly speaking, the correlation operator does not have
eigenvectors \emph{in} the Hilbert space. The reason for this, is
that this operator $C$, as it also happens with position and
momentum operators, is unbound as can be proven from its definition
in Eq.(\ref{Corr}). If we assume the existence of the eigenvectors
$\{\eta_{c}\}$ we can arrive at several contradictions. One of them
arises from the commutation relations of the type $[B,C]=ikB$ shown
in Eqs.(\ref{ComRel} -\ref{ComRel4}). This commutation relation
implies that the operator $B$ acts as a ``shift'' operator for the
eigenvectors of $C$. In fact, it can be easily shown that if
$\eta_{c}$ is an eigenvector of $C$ with eigenvalue $c$, then
$B\eta_{c}$ is also an eigenvector corresponding to the eigenvalue
$(c-ik)$; but then the \emph{hermitian} operator $C$ could have
\emph{complex} eigenvalues reaching a contradiction.

In quantum mechanics, however, we need the eigenvectors of unbound
operators like $X$, $P$ or $C$ because they represent possible
states of the system. There are two standard ways out of this
difficulty. One of them is to assume for the Hilbert space, not the
squared integrable \emph{functions} but instead,
\emph{distributions} that include also non squared integrable
functions, like Dirac's delta ``function'' and $e^{ikx}$. These are
precisely the eigenvectors of $X$ and $P$. The other way,
mathematically more elegant, is to consider the Gel'fand triplet
$\mathcal{H}^{0}\subseteq\mathcal{H}\subseteq\mathcal{H}'$ that
amounts to an extension of the Hilbert  space $\mathcal{H}$ towards
the so-called \emph{Rigged} Hilbert space $\mathcal{H}'$ that
includes the desired eigenvectors\cite{bal}.

We have then the three bases $\{\varphi_{x}\},\ \{\phi_{p}\}$ and
$\{\eta_{c}\}$ associated to $X, P$ and $C$. The transformation
between the first two, already known, is given by
$\langle\varphi_{x},\phi_{p}\rangle=
\frac{1}{\sqrt{2\pi}}\exp(\frac{i}{\hbar}px)$ and we must now
determine $\eta_{c}(x)=\langle\varphi_{x},\eta_{c}\rangle$ and
$\eta_{c}(p)=\langle \phi_{p},\eta_{c}\rangle$, that is, the
eigenvectors of $C$ in position and momentum representation. For
this, we can write the eigenvalue Eq.(\ref{eigenvec}) in the
position or momentum representation and solve it to find the
associated eigenfunctions. That is, we must solve
\begin{eqnarray}
 x\frac{d\eta_{c}(x)}{dx}&=&
  \left(i\frac{c}{\hbar}-\frac{1}{2}\right)\eta_{c}(x) \\
 p\frac{d\eta_{c}(p)}{dp}&=&
 \left(-i\frac{c}{\hbar}-\frac{1}{2}\right)\eta_{c}(p)\ .
\end{eqnarray}
(To avoid confusion, notice that we are using the same letter,
$\eta$, to denote \emph{different} functions $\eta_{c}(x)$ and
$\eta_{c}(p)$ that are actually related by Fourier transformation).

The correlation operator is invariant under the (unitary) parity
transformation ${\mathcal{P}}$ that changes $X\rightarrow-X$ and
$P\rightarrow-P$. That is, $[C,\mathcal{P}]=0$ and this implies that
the eigenvalues are also invariant. That is, if $\eta_{c}$ is an
eigenvector, then, $\mathcal{P}\eta_{c}$ is also an eigenvector with
the same eigenvalue and therefore the correlation eigenvalues are
twofold degenerate because the parity operator has two eigenvectors,
even (\emph{gerade}) $\eta^{g}_{c}$ or odd (\emph{ungerade})
$\eta^{u}_{c}$. These two eigenvectors are orthogonal because
$\left\langle\eta^{g}_{c},\eta^{u}_{c}\right\rangle =
\left\langle\eta^{g}_{c},\mathcal{P}^{2}\eta^{u}_{c}\right\rangle =
\left\langle\mathcal{P}^{\dag}\eta^{g}_{c},\mathcal{P}\eta^{u}_{c}\right\rangle
= \left\langle\eta^{g}_{c},-\eta^{u}_{c}\right\rangle =
-\left\langle\eta^{g}_{c},\eta^{u}_{c}\right\rangle$. The explicit
treatment of the above equation in the position representation
provides both degenerate solutions:
\begin{equation}\label{eqn:corr2g}
 \eta^{g}_{c} (x)=\frac{1}{2\sqrt{\hbar\pi}}\ |x|^{-\frac{1}{2}+i\frac{c}{\hbar}}
 =\frac{1}{2\sqrt{\hbar\pi}}\frac{e^{i\frac{c}{\hbar}\ln|x|}}{\sqrt{|x|}} \ ,
\end{equation}
\begin{equation}\label{eqn:corr2u}
 \eta^{u}_{c} (x)=\frac{\mathrm{sign}(x)}{2\sqrt{\hbar\pi}}\ |x|^{-\frac{1}{2}+i\frac{c}{\hbar}}
 =\frac{\mathrm{sign}(x)}{2\sqrt{\hbar\pi}}\frac{e^{i\frac{c}{\hbar}\ln|x|}}{\sqrt{|x|}} \ ,
\end{equation}
normalized such that
$\langle\eta^{k}_{c},\eta^{k'}_{c'}\rangle=\delta_{k,k'}\
\delta(c-c')$. The momentum representation of the eigenfunctions can
be obtained in the same way, that is, solving Eq.(17), or by taking
the Fourier transform of Eqs.(\ref{eqn:corr2g},\ref{eqn:corr2u}) or,
most easily, by noticing that the operator $C$ in the momentum
representation is obtained from the position representation by
replacing $x\rightarrow p$ and taking the complex conjugate.
Therefore, if $\eta_{c}(x)$ is an eigenfunction in the position
representation, then $\eta_{c}^{\ast}(p)$ is the corresponding
eigenfunction in the momentum representation. These eigenfunctions
have the interesting property that their Fourier transformation is
equal to their complex conjugate.

\section{PAULI PARTNERS AMBIGUITY }
We can now analyse the possibility to resolve the Pauli partner
ambiguity by means of the correlation operator $C$. Let us recall
that two \emph{different} states $\Psi$ and $\Phi$ (that is,
$|\langle\Psi,\Phi\rangle|\lneqq 1 $) are Pauli partners if they
have equal position and momentum existential weights $\rho (x)$ and
$\varpi (p)$. That is,
$|\langle\varphi_{x},\Psi\rangle|=|\langle\varphi_{x},\Phi\rangle|$
and $|\langle\phi_{p},\Psi\rangle|=|\langle\phi_{p},\Phi\rangle|$.
This means that the states in position representation
$\langle\varphi_{x},\Psi\rangle$ and
$\langle\varphi_{x},\Phi\rangle$ differ at most by a phase
$e^{i\alpha(x)}$. This condition in the abstract Hilbert space is
that there exists an (hermitian) operator function of position
$\alpha(X)$ such that
\begin{equation}\label{alf }
   \Psi = e^{i\alpha(X)}\Phi\ ,
\end{equation}
and from the equality of the momentum existential weight it follows
that there exist an operator function $\beta(P)$ such that
\begin{equation}\label{bet }
   \Psi = e^{i\beta(P)}\Phi\ .
\end{equation}
%\begin{eqnarray}
% \nonumber to remove numbering (before each equation)
%\nonumber   &e^{i\alpha(X)}&   \\
%\nonumber \Phi \bullet &\Longrightarrow& \bullet \Psi \\
%\nonumber   &e^{i\beta(P)}&
%\end{eqnarray}
The Pauli partners $\Psi$ and $\Phi$ are therefore eigenstates, or
fix points, of two unitary operators,
\begin{eqnarray}
% \nonumber to remove numbering (before each equation)
  e^{i\alpha(X)}e^{-i\beta(P)}\Psi&=& \Psi\\
  e^{-i\alpha(X)}e^{i\beta(P)}\Phi&=& \Phi \ .
\end{eqnarray}
To avoid misunderstanding it must be clear that the two operators
$e^{i\alpha(X)}$ and $e^{i\beta(P)}$ are different operators but for
the pair $(\Psi,\Phi)$ they have the same effect, that is,
$\Phi\rightarrow\Psi$. For any other Hilbert space element they
produce different results; clearly, we can \emph{not} represent
these operators by $\Psi\langle\Phi,\cdot\rangle$. The main
difficulty in dealing with Pauli partners is that we know that they
exist, but we do not have a complete characterization of them.
Therefore we can not give explicit expressions for the functions
$\alpha(X)$ and $\beta(P)$ and, for instance, we don't know the
commutator $[\alpha(X),\beta(P)]$.

In a numerical survey\cite{goye} done with an iterative algorithm
for the determination of states in finite dimensional Hilbert
spaces\cite{dlt4}, large number of Pauli partners were found and
\emph{in all cases} the partners were differentiated by the
correlation operator.  With this numerical result one could jump to
the conjecture that the correlation observable always resolves the
Pauli partner ambiguity. However this conclusion could be wrong
because the existence of a set of Pauli partners with null measure
is not excluded in a numerical survey and therefore we can never be
sure to have analysed all Pauli parters. Furthermore the survey
involves only low dimensional Hilbert spaces and there is no
guaranty that the same is true in infinite dimensions.

For an analytical treatment of the possibility to resolve the Pauli
partner ambiguity by means of the correlation observable, we should
calculate and compare the two distributions
$\sigma_{\Psi}(c)=|\langle\eta^{g}_{c},\Psi\rangle|^{2}+|\langle\eta^{u}_{c},\Psi\rangle|^{2}$
and
$\sigma_{\Phi}(c)=|\langle\eta^{g}_{c},\Phi\rangle|^{2}+|\langle\eta^{u}_{c},\Phi\rangle|^{2}$.
A simpler approach, motivated by the numerical survey, is to look at
the expectation values of theses distributions and compare
$\langle\Psi,C\Psi\rangle$ with $\langle\Phi,C\Phi\rangle$. This is
however also affected by the same difficulties mentioned before and
the question whether the additional information provided by the
correlation observable is sufficient in order to solve the Pauli
ambiguity is still open.

\section{CORRELATION IN THE QFT INTERPRETATION OF QM}

Position-momentum correlations have a simple explanation in an
interpretation of quantum mechanics (QM) suggested by quantum field
theory (QFT). In this interpretation, consistent with the
ontological choice for the indeterminacies mentioned in fourth
section of this work,  we can view the ``probability cloud'' as a
permanent creation, propagation and annihilation of virtual
particles in an indefinite number making up the quantum field
associated to some particle type. We can think that the virtual
particles are the components of the field that have objective but
ephemeral existence with position and momentum. In this view, the
Feynman graphs are not only mathematical terms of a perturbation
expansion but represent real excitations of the quantum field.

Let us imagine then virtual components of the field created at a
location at ``the right'' of the one dimensional distribution
$\rho(x) $, that is with a \emph{positive} value for the observable
$X-\langle X\rangle$. If these components are moving with momentum
smaller than the mean value, that is, with \emph{negative} value for
$P-\langle P\rangle$ the relative motion will be towards the center
and the distribution will shrink. Similarly, the components created
at the left and moving to the right have the two offsets $X-\langle
X\rangle$ and $P-\langle P\rangle$ with different sign, that is,
their (symmetrized) product is negative.

For simplicity, let us assume that in this state we have $ \langle
X\rangle=\langle P\rangle =0$ (the general state is obtained with
the translation and impulsion operator). Therefore the product of
the two offsets in position and momentum is precisely the
correlation observable and the previous argument means that if the
correlation is negative the space distribution shrinks. We can prove
this with rigour: let us calculate the time derivative of the width
of the distribution $\Delta^{2} x = \langle X^{2}\rangle $. In the
Heisenberg picture, assuming a nonrelativistic hamiltonian
$H=P^{2}/2m$, we have
\begin{equation}\label{shrink}
 \frac{dX^{2}}{dt}=\frac{-i}{\hbar}[X^{2},H]=\frac{-i}{2\hbar m}[X^{2},P^{2}]
 =\frac{1}{m}(XP+PX)=\frac{2}{ m}C.
\end{equation}
Taking expectation values we conclude that states with negative
correlation shrink and states with positive correlation expand, as
expected from the heuristic argument given above.

The momentum distribution for a free particle is time independent
and if the state is shrinking, that is, with negative correlation,
we are approaching the limit imposed by Heisenberg indeterminacy
principle. This principle will not be violated because the
correlation will not remain always negative: at some time it will
become positive and the state will begin to expand. In fact, we can
prove that the correlation is always increasing in time:
\begin{equation}\label{corrtimeincr}
 \frac{dC}{dt}=\frac{-i}{\hbar}[C,H]=\frac{-i}{4\hbar m}[XP+PX,P^{2}]
 =\frac{1}{m}P^{2}=2H,
\end{equation}
and this is a nonnegative operator. If a state is shrinking, at some
later time it will be spreading. Gaussian states of this sort have
been reported\cite{dlt4} in a very comprehensive paper.

It is interesting to notice that the fact that the correlation (like
entropy in thermodynamics) is always increasing can be used to
define a \emph{quantum mechanical arrow of time} without recourse to
the state collapse that is one of the most mysterious features of
quantum mechanics.
\section{FINAL COMMENTS}
In this work we have seen that position and momentum observables in
quantum mechanics are more subtle than their corresponding classical
variables. In particular it is interesting to notice that these are
the unique relevant observables for the simplest quantum system of a
free particle, but they are \emph{not sufficient} to fix the state:
in this quantum system there must be something else that has to be
specified for an unambiguous determination of its behaviour. It is
suggested that the correlation between position and momentum can
play this role.

We have presented several features of the correlation that becomes
an intuitive explanation in an interpretation of quantum mechanics
where the virtual particles acquire real, but ephemeral, existence.
It is remarkable that the sign of the correlation expectation value
controls the shrinking and spreading of a wave packet. The
correlation for a free particle is always increasing and therefore
in the long term states expand. In this context, it is easy to prove
that there is an important class of states --the coherent states--
with vanishing expectation value for the correlation (but there are
other states, not coherent, that also have zero expectation value
for the correlation).

Another place where the correlation appears, not shown in this work
but related with the comment above, is in the improved version of
Heisenberg´s uncertainty principle, derived by
Schr\"{o}dinger\cite{sch,dlt5}, where an extra term involving the
anti-commutator (that is, the correlation) besides the commutator
contribution limit the uncertainty product.

Finally let us conclude with some speculative comments. We usually
try to understand quantum mechanics as an extension of classical
mechanics with new concepts beyond classical physics. So, to the
energy of an oscillator we must add the \emph{zero point energy}
$\omega\hbar/2$ and to the orbital angular momentum we include the
\emph{intrinsic spin} $\hbar/2$ that have an essential quantum
origin. With the correlation something similar happens: if we use
the position-momentum commutator to write it as $C=PX+i\hbar/2$, we
see that an essential quantum contribution $\hbar/2$ is added to the
classical correlation. If we are ever to have a different paradigm
to explain quantum mechanics, it will have to bring some rationale
for the zero point energy, the zero point angular momentum and the
zero point correlation.

\end{document}